\documentclass[12pt]{article}
\usepackage{graphicx}

\newcommand{\Frac}[2]{\frac{\displaystyle #1}{\displaystyle #2}}
\newcommand{\beq}{\begin{equation}}
\newcommand{\eeq}{\end{equation}}
\newcommand{\beqn}{\begin{eqnarray}}
\newcommand{\eeqn}{\end{eqnarray}}
\newcommand{\beqns}{\begin{eqnarray*}}
\newcommand{\eeqns}{\end{eqnarray*}}

\begin{document}

\begin{titlepage}
\begin{center}

\hfill USTC-ICTS/PCFT-25-60\\
\hfill December 2025

\vspace{2.5cm}
{\large {\bf  Improved analysis of rare $Z$-boson decays into a heavy vector quarkonium plus lepton pair}} \vspace*{1.0cm}\\
{ Li Ang$^\dagger$ and Dao-Neng Gao$^\ddagger$} \vspace*{0.3cm} \\
{\it\small
Interdisciplinary Center for Theoretical Study, University of Science and Technology of China,
Hefei, Anhui 230026 China}\\
{\it\small Peng Huanwu Center for Fundamental Theory, Hefei, Anhui 230026 China}
\vspace*{1cm}

\end{center}
\begin{abstract}
\noindent
 We improve the theoretical predictions for rare $Z$-boson decays, $Z\to V\ell^+\ell^-$ ($\ell=e$ or $\mu$), where $V$ denotes a heavy vector quarkonium including $J/\Psi$, $\Psi (2S)$, and $\Upsilon (nS)$ with $n=1,2,3$. These processes are thought to be dominated by the electromagnetic fragmentation transition, i.e.,  $Z\to \gamma^*\ell^+\ell^-$ followed by $\gamma^*\rightarrow V$. The present study includes all of the relevant tree-level Feynman diagrams, which contribute to these decays in the standard model. Our analysis shows that, for the charmonium final states, the fragmentation transition almost saturates the whole contribution and the other diagrams can be neglected; while for the bottomonium final states, the inclusion of other diagrams can increase their branching fractions by $4\%\sim 9\%$. Further investigation of the differential distributions, especially the angular distributions, indicates that forward-backward asymmetries for final leptons in these processes would be zero in the standard model. Therefore, in future experimental facilities with large number of $Z$-boson events accumulated, studies of these rare $Z$ decays may help both to test the standard model and to probe its interesting extensions.

\end{abstract}

\vfill
\noindent
$^{\dagger}$ E-mail address:~tamaki@mail.ustc.edu.cn\\\noindent
$^{\ddagger}$ E-mail address:~gaodn@ustc.edu.cn

\end{titlepage}
\vspace{0.5cm}

\section{Introduction}

The experimental discovery of the Higgs boson at the Large Hadron Collider (LHC) \cite{atlascms}, which completes the full particle content predicted by the standard model (SM), has opened up a new era of precision measurements in particle physics. Of all the existing elementary particles, the top quark, the Higgs boson, and the intermediate weak gauge bosons ($W$ and $Z$) are the most massive ones, with masses around the electroweak scale $\sim {\cal O}(100)$ GeV. Thus the in-depth investigations, both experimentally and theoretically, of the properties of these heavy elementary particles will play an important role in precision SM tests and in looking for new physics beyond the SM.

Although the $W$ and $Z$ bosons were discovered more than forty years ago, only exclusive leptonic decays of theirs were observed, as shown by Particle Data Group \cite{PDG24},  and no exclusive hadronic decays have been well established so far. It is of interest to study the exclusive gauge boson decaying into hadronic final states, which may present us with a new playground for precision electroweak and quantum chromodynamics (QCD) physics \cite{GKPR80,AMP82,GKN15, EL25}.

Several years ago, the CMS Collaboration reported the first observation of a rare $Z$-boson three-body decay, $Z\to \Psi \ell^+\ell^-$ ($\ell=e$ or $\mu$) \cite{CMS18}, where $\Psi$ represents direct $J/\Psi$ and $J/\Psi$ from $\Psi(2S)\to J/\Psi X$ decays. After subtraction of the contribution from $\Psi(2S)$, the branching fraction ratio
\beq\label{RofZtoJpsi}
{\cal R}_{J/\Psi\ell^+\ell^-}=\frac{{\cal B}(Z\to J/\Psi\ell^+\ell^-)}{{\cal B}(Z\to \mu^+\mu^-\mu^+\mu^-)}=0.67\pm 0.18({\rm stat})\pm 0.05({\rm syst})
\eeq
has been obtained, and using the measured value of ${\cal B}(Z\to \mu^+\mu^-\mu^+\mu^-)$ given in Ref. \cite{CMS18-2}, they estimated that ${\cal B}(Z\to J/\Psi\ell^+\ell^-)$ could be about $8\times 10^{-7}$.

On the other hand, one may expect that, to search for exclusive hadronic $Z$-boson decays, the two-body radiative decays $Z\to V\gamma$ with $V$ denoting a neutral vector quarkonium like $J/\Psi$ or $\Upsilon$ etc, should be more promising than the above three-body processes. Naively, the $Z\to V\ell^+\ell^-$ transition is suppressed by a power of $\alpha_{\rm em}$, where $\alpha_{\rm em}$ is the electromagnetic coupling constant, compared with the radiative decay. The experimental searches for $Z$ decays into a heavy vector quarkonium plus a photon have been performed by the ATLAS and CMS Collaborations at the LHC \cite{ZVgamma-LHC}, but so far, no significant signals have been observed. From a theoretical perspective, $Z\to V\gamma$ decays have been extensively investigated \cite{GKPR80,GKN15,EL25, HP15, BCEL18,ZVgamma-theory} and their branching ratios are predicted to be around $10^{-8}$ or even smaller, which is obviously smaller than the value of ${\cal B}(Z\to J/\Psi \ell^+\ell^-)$ given by eq. (\ref{RofZtoJpsi}).
\begin{figure}[t]
\begin{center}
\includegraphics[width=6cm,height=3cm]{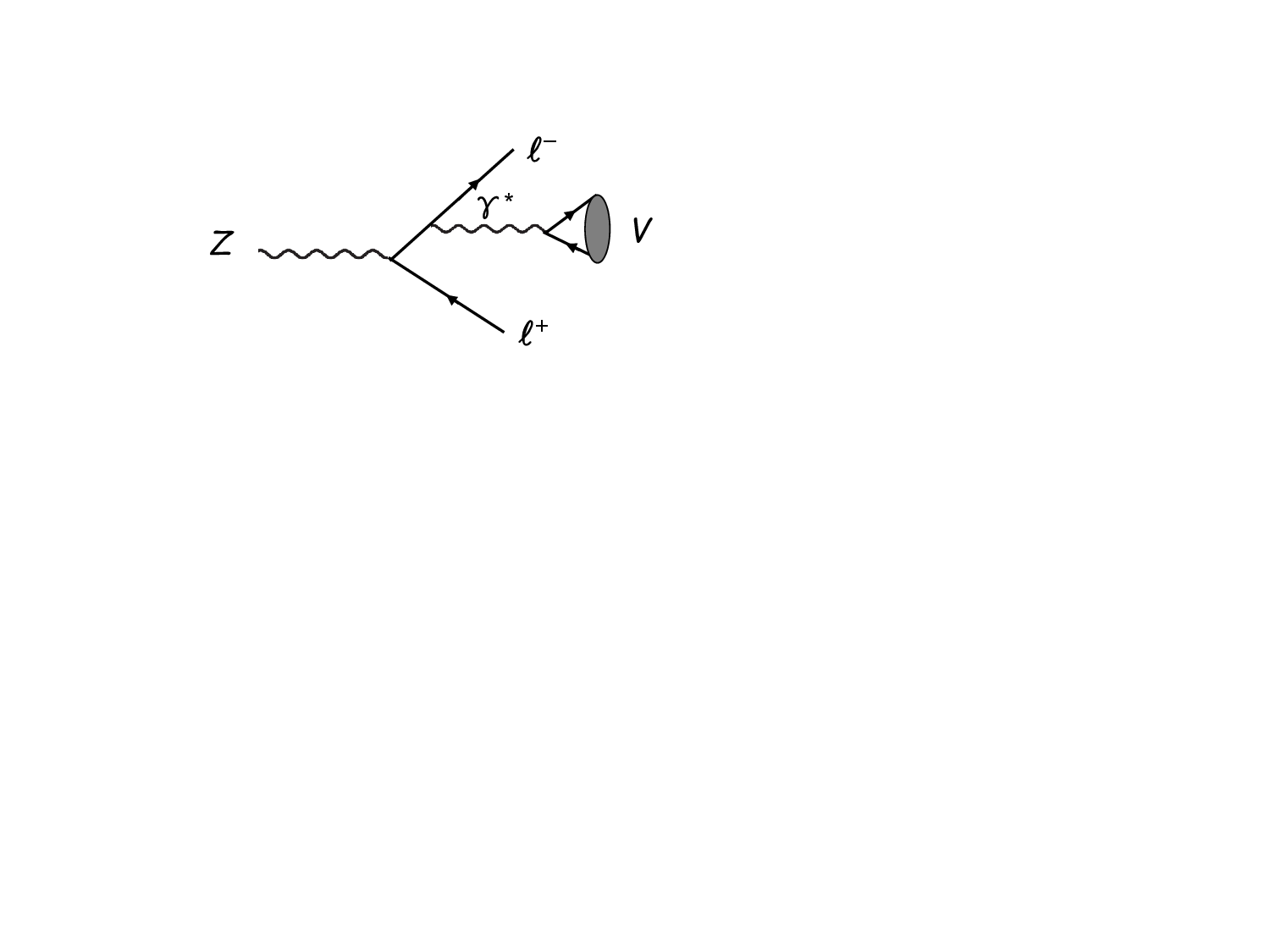}
\end{center}
\caption{The dominant Feynman diagrams for $Z\to V \ell^+ \ell^-$ decays. The virtual photon $\gamma^*$ could also be emitted from the $\ell^+$ line.}\label{figure1}
\end{figure}

Actually, theoretical analysis of $Z\to V\ell^+\ell^-$ decays has been carried out and it has been pointed out in Refs. \cite{BR90, Fleming93, Fleming94} that the dominant contribution to these processes in the SM comes from $Z\to \gamma^* \ell^+\ell^-$ with the subsequent transition $\gamma^*\to V$. The corresponding Feynman diagrams have been depicted in Fig. \ref{figure1}, and the branching ratios of these decays have been calculated. For the case where $V=J/\Psi$, ${\cal B}(Z\to J/\Psi\ell^+\ell^-)$ has been estimated as $(6.7\pm 0.7)\times 10^{-7}$ \cite{BR90} and $7.5\times 10^{-7}$ \cite{Fleming94}, respectively, which are consistent with the present experimental observation.
As explained in Refs. \cite{Fleming93, Fleming94}, the unexpectedly large decay rates of $Z\to V\ell^+\ell^-$, contributed by Fig. \ref{figure1}, could be understood due to an electromagnetic fragmentation transition, $\gamma^*\to V$,  which is not suppressed by a factor $m_V^2/m_Z^2$, compared with those of $Z\to V\gamma$ transitions. This indicates that these three-body $Z$ decays may be very promising candidates in future experimental facilities, where large number of $Z$ bosons can be produced.

In Refs. \cite{BR90, Fleming93, Fleming94}, only the dominant electromagnetic fragmentation contribution from Fig. \ref{figure1} was considered, and the author of Refs. \cite{Fleming93, Fleming94} has argued that the other contributions, for instance, from $Z\to V\gamma^*\to V\ell^+\ell^-$, shown in Fig. \ref{figure2}, will be suppressed by $m_V^2/m_Z^2$. It is easy to see that, for the charmonium resonances, the factor $m_\Psi^2/m_Z^2$ is about $10^{-3}$, the electromagnetic fragmentation transition  may almost saturate the whole contribution of $Z\to J/\Psi \ell^+\ell^-$; while for the bottomonium case, other contributions might bring about the effects at the percent level. Note that the amplitude from Fig. \ref{figure2} is at the same order in the coupling constants as that from Fig. \ref{figure1}. Therefore, it is of interest to perform a systematical calculation of the whole tree-level SM contributions to $Z\to V\ell^+\ell^-$ decays, which will help to compare the SM predictions with future precise measurements. Meanwhile, in addition to the decay rates, the three-body processes could generate some significant angular information, which may induce some interesting observable like the forward-backward asymmetries of leptons in future high-precise experimental facilities.

The remainder of this paper is organized as follows. In Section 2, we present a detailed derivation of the tree-level decay amplitudes in the SM. Section 3 is our phenomenological analysis, both in the SM and beyond the SM. We summarize our results and give some outlook in Section 4.

\section{Decay amplitudes}

\begin{figure}[t]
\begin{center}
\includegraphics[width=6cm,height=3cm]{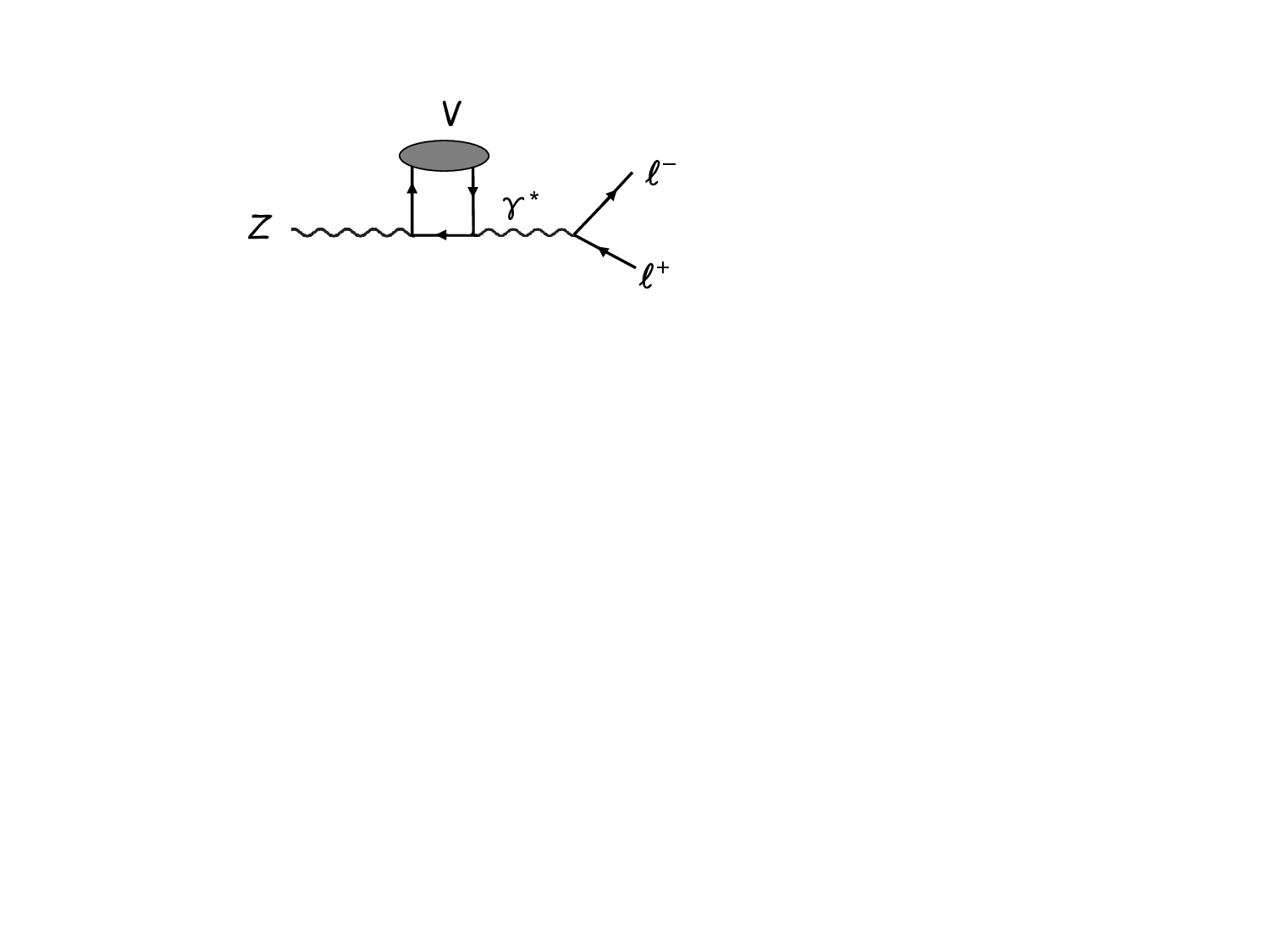}
\end{center}
\caption{The Feynman diagram contributing to $Z\to V\ell^+\ell^-$ via $Z\to V\gamma^*$ transitions. The virtual photon $\gamma^*$ could also be emitted from the left quark line.}\label{figure2}
\end{figure}

 From Figs. \ref{figure1} and \ref{figure2}, one can easily find that these diagrams contain the standard neutral current interactions
\beq\label{NC}
{\cal L}_{\rm NC}=e J_\mu^{\rm em} A^\mu+\frac{g_Z}{2}J^Z_\mu Z^\mu\eeq
with
\beq\label{em-weak-current}
J_\mu^{\rm em}=\sum_f Q_f\bar{f}\gamma_\mu f,\;\;\;J_\mu^Z=\sum_f \bar{f}\gamma_\mu(g_V^f-g_A^f\gamma_5) f.
\eeq
Here $f$ denotes fermions including leptons ($\ell$) and quarks ($Q$), $e$ is the QED coupling constant, $Q_f$ is the electric charge, and $g_Z$ denotes the overall coupling strength of the $Z$-boson to fermions. We write
\beqn\label{gZ}
&&g_Z=2 (\sqrt{2}G_F)^{\frac{1}{2}} m_Z,\nonumber\\
&& g_V^f=T_3^f-2 Q_f \sin^2 \theta_W,\;\;\;\;g_A^f=T_3^f,
\eeqn
where $G_F$ is the Fermi constant, $\theta_W$ is the Weinberg angle, and $T_3^f$ is the third component of the weak isospin of the fermion.

Thus one can directly derive the decay amplitude for Fig. \ref{figure1} as
\beqn\label{amp0}
i{\cal M}_1^\gamma&=&\frac{ie^2g_Z Q_V f_V}{2 m_V}\epsilon_\nu(p)\epsilon_\mu^*(q)\bar{u}(k_1)\left(\gamma^\mu \frac{k_1\!\!\!\!\!/+q\!\!\!/+m_\ell}{m_V^2+2k_1\cdot q}\gamma^\nu
-\gamma^\nu \frac{k_2\!\!\!\!\!/+q\!\!\!/+m_\ell}{m_V^2+2k_2\cdot q}\gamma^\mu\right)\nonumber\\ &&\times (g_V^\ell-g_A^\ell\gamma_5)v(k_2),
\eeqn
where $p$, $q$, $k_1$, and $k_2$ represent the momenta of $Z$, $V$, $\ell^-$, and $\ell^+$, respectively. $Q_V$ is the electric charge of the quark inside $V$. $f_V$ is the decay constant of $V$, which is defined by
 \beq\label{fv}
\langle V(p,\epsilon)|\bar{Q}\gamma_\mu Q |0\rangle= f_{V}m_{V}\epsilon^{*}_\mu.
\eeq
Here $\epsilon^{*}_\mu$ is polarization vector of $V$, and the value of $f_V$ can be fixed from the measured $V\to e^+e^-$ width through
\beq\label{vee}
\Gamma(V\to e^+ e^-)=\frac{4\pi Q_V^2 f_V^2}{3m_V}\alpha^2_{\rm em}(m_V).
\eeq
Note that the definition (\ref{fv}) also fulfills the hadronization of the electromagnetic current $\bar{Q}\gamma_\mu Q$ into the final particle $V$.

Different from the case of ${\cal M}_1^\gamma$, one can find that, in Fig. \ref{figure2}, $Z$-boson first couples to a heavy quark pair $Q\bar{Q}$, then converts into the final lepton pair via the virtual photon intermediate state, which is emitted by one of the heavy quark lines. In order to obtain the hadronic decay amplitudes, one has to project $Q\bar{Q}$ into the corresponding heavy quarkonium states. As a reasonable approximation for the leading order calculation,  in this work we will adopt the nonrelativistic color-singlet model \cite{colorsinglet}, in which the quark momentum and mass are taken to be one half of the corresponding quarkonium momentum $q$ and mass $m_V$, namely, $q_Q=q_{\bar{Q}}=q/2$ and $m_{V}=2m_Q$.  According to Refs. \cite{barger, Jia2007}, to form the heavy quarkonium $V$, one can thus replace the combination of the Dirac spinors of heavy quarks by the following projection operator
\beq\label{projector}
{v(q_{\bar{Q}})} {\bar{u}(q_Q)} \longrightarrow \frac{\psi_{V}(0)I_c}{2\sqrt{3m_V}} {\epsilon\!\!/}^*(q\!\!\!/+m_V),
\eeq
where $I_c$ is the $3\times 3$ unit matrix in color space and $\epsilon^{*\mu}$ is the polarization vector of the quarkonium $V$. $\psi_{V}(0)$ represents the wave function at the origin for $V$, which is a nonperturbative quantity. Now from Fig. \ref{figure2},  it is straightforward to get the amplitude as
\beqn\label{amp2}
i{\cal M}_2=4\sqrt{3}e^2g_Z g_A^Q m_V\left(\frac{\psi_V(0)}{\sqrt{m_V}}\right)\varepsilon^{\mu\nu\alpha\beta}\epsilon_\nu(p)\epsilon_\mu^*(q)k_\alpha \frac{ \bar{u}(k_1)\gamma_\beta v(k_2)}{(m_Z^2-m_V^2+k^2)k^2},
\eeqn
where $k=k_1+k_2$.

It is obvious that both ${\cal M}_1^\gamma$ and ${\cal M}_2$ are ${\cal O}(e^2 g_Z)$. Actually, if we replace the virtual photon in Figs. \ref{figure1} and \ref{figure2} by a virtual $Z$-boson, the corresponding amplitudes are ${\cal O} (g_Z^3)$, which are of the same order in electroweak coupling constants as ${\cal M}_1^\gamma$ and ${\cal M}_2$.  It will be shown below that, these ${\cal O} (g_Z^3)$ amplitudes do not bring about significant contributions. However, in order to perform a complete calculation, we include both of them in the present study. It is easy to see that the amplitude contributed by $Z\to \ell^+\ell^- Z^*$, followed by $Z^*\to V$, can be incorporated into ${\cal M}_1^\gamma$, and their sum can be written in a compact form, which, denoted as ${\cal M}_1$, gives \footnote{Now ${\cal M}_1$ looks like ${\cal O}(e^2 g_Z)$. However, the $Z^*\to V$ amplitude can be restored as ${\cal O} (g_Z^3)$  by using the relation $e^2/\sin^2 2\theta_W\sim g_Z^2$ in the second term of $\tilde{g}_V^\ell$ and $\tilde{g}_A^\ell$ in eq. (\ref{gvatilde}).}
\beqn\label{amp1}
i{\cal M}_1&=&\frac{ie^2g_Z Q_V f_V}{2m_V}\epsilon_\nu(p)\epsilon_\mu^*(q)\bar{u}(k_1)\left(\gamma^\mu \frac{k_1\!\!\!\!\!/+q\!\!\!/+m_\ell}{m_V^2+2k_1\cdot q}\gamma^\nu
-\gamma^\nu \frac{k_2\!\!\!\!\!/+q\!\!\!/+m_\ell}{m_V^2+2k_2\cdot q}\gamma^\mu\right)\nonumber\\ &&\times (\tilde{g}_V^\ell-\tilde{g}_A^\ell\gamma_5)v(k_2)
\eeqn
with
\beqn\label{gvatilde}
\tilde{g}_V^\ell=g_V^\ell+\frac{m_V^2[(g_V^\ell)^2+(g_A^\ell)^2] g_V^Q}{\sin^2 2 \theta_W(m_Z^2-m_V^2)Q_V},\;\;\;\; \tilde{g}_A^\ell=g_A^\ell+\frac{m_V^2 (2 g_V^\ell g_A^\ell) g_V^Q}{\sin^2 2 \theta_W(m_Z^2-m_V^2)Q_V}.
\eeqn
 Since the axial part in $J_\mu^Z$ of eq. (\ref{em-weak-current}) does not contribute to $Z^*\to V$, only $g_V^Q$ appears in eq. (\ref{gvatilde}).  Also, by replacing $\gamma^*$ with $Z^*$ in Fig. \ref{figure2}, one can similarly obtain
\beqn\label{amp3-Z}
i{\cal M}_3=-\sqrt{3}g_Z^3 g_V^Q g_A^Q m_V\left(\frac{\psi_V(0)}{\sqrt{m_V}}\right)\varepsilon^{\mu\nu\alpha\beta}\epsilon_\nu(p)\epsilon_\mu^*(q)(2k+q)_\alpha \frac{\bar{u}(k_1)\gamma_\beta(g_V^\ell-g_A^\ell \gamma_5) v(k_2)}{(m_Z^2-m_V^2+k^2)(k^2-m_Z^2)}.
\eeqn
Therefore, the full tree-level decay amplitude of $Z\to V\ell^+\ell^-$ in the SM reads
\beq\label{amp-tot}
{\cal M}={\cal M}_1+{\cal M}_2+{\cal M}_3.\eeq

After squaring the amplitude and summing or averaging over the polarizations of final of initial particles, the differential decay rate of $Z\to V \ell^+\ell^-$ can be obtained as
\beq\label{distribution1}
\frac{d^2\Gamma}{d s ~d\cos\theta}=\frac{1}{512\pi^3m_Z^3}\beta_\ell ~\lambda^{1/2}(m_Z^2, m_V^2,s)~ \frac{1}{3}\sum_{\rm spins}|{\cal M}|^2
\eeq
 with $\beta_\ell=\sqrt{1-4 m_\ell^2/s}$, $\lambda(a,b,c)=a^2+b^2+c^2-2ab-2ac-2bc$, and $s=k^2$. Here $\theta$ is the angle between the three-momentum of $Z$-boson and the three-momentum of $\ell^-$ in the dilepton rest frame, and the phase space is given by
\beq\label{phasespace}
4 m_\ell^2\leq s \leq (m_Z-m_V)^2,\;\;\;\; -1\leq\cos\theta\leq 1.
\eeq
By expressing the decay rate as
\beq\label{rate}
\Gamma(Z\to V\ell^+{\ell}^-)=\Gamma_1+\Gamma_2+\Gamma_3+\Gamma_{12}+\Gamma_{13}+\Gamma_{23},
\eeq
together with the total decay width of $Z$-boson $\Gamma_Z=2.4955$ GeV \cite{PDG24}, we can define the branching ratios as
\beq\label{branchingratioi}
{\cal B}_i(Z\to V\ell^+\ell^-)={\Gamma_i}/{\Gamma_Z}
\eeq
and
\beq\label{branchingratio}
{\cal B}(Z\to V \ell^+\ell^-)={\Gamma(Z\to V \ell^+\ell^-)}/{\Gamma_Z}.
\eeq
Here $\Gamma_i$ denotes the contribution from ${\cal M}_i$ for $i=1,2,3$, respectively, and $\Gamma_{ij}$ is due to the interference between ${\cal M}_i$ and ${\cal M}_j$. The lepton mass can be neglected in calculating the amplitudes squared; while we retain nonzero $m_\ell$ for kinematics in order to avoid the infrared divergence in $\Gamma_2$.

On the other hand, the analysis of the differential distributions with respect to $s$ or $\cos\theta$ will be also significant, which can be carried out by integrating over $\cos\theta$ or $s$ in eq. (\ref{distribution1}).  Particularly, as shown in the Appendix, the differential rate $d^2\Gamma_1/ds~d\cos\theta$ is the even function of $\cos\theta$, and the same conclusion can be reached for other differential rates $d^2\Gamma_i/ds~d\cos\theta$ with $i\neq 1$. Thus, the angular distribution is symmetric under $\cos\theta\leftrightarrow -\cos\theta$, consequently, no forward-backward asymmetries for final leptons can be expected in the SM. This implies that the forward-backward asymmetry, which is defined as
\beqn\label{AFB1}
A_{\rm FB}(s)=\Frac{\int_0^1 \frac{d^2\Gamma }{d s ~d\cos\theta} d\cos\theta-\int_{-1}^0 \frac{d^2\Gamma }{{d s}~{d\cos\theta}} d{\cos\theta}}{\int_0^1 \frac{{d^2\Gamma}}{{d s}~{d\cos\theta}}  d{\cos\theta}+\int_{-1}^0 \frac{{d^2\Gamma }}{{d s}~{d\cos\theta}}d{\cos\theta}},
\eeqn
could be an interesting observable to probe new physics beyond the SM. In the next section, we will analyze this quantity by introducing the anomalous neutral gauge coupling terms. It will be shown below, due to the new interaction, the linear term of $\cos\theta$ could appear in the differential distributions, which may lead to nonzero $A_{\rm FB}$ in $Z\to V\ell^+\ell^-$ decays.

\section{Phenomenological analysis}

We are now ready to perform the numerical studies of $Z\to V\ell^+\ell^-$ decays with $\ell=e,~\mu$. At first, we present detailed predictions on branching fractions and differential distributions for these processes in the SM. In the second part, we discuss some possible new physics effects in these decays.

\subsection{In the SM}

By using the experimental data of $\Gamma(J/\Psi\to e^+e^-)=(5.53\pm 0.10)$ keV given by Particle Data Group \cite{PDG24}, and taking $\psi^2_{J/\Psi}(0)=0.073^{+0.011}_{-0.009}$ GeV$^3$ from Ref. \cite{BCKLY08}, for $Z\to J/\Psi e^+ e^-$, one will get
\beqn\label{brJpsi}
&&{\cal B}_1=7.78\times 10^{-7},\;\;\; {\cal B}_2=1.62\times 10^{-9},\;\;\; {\cal B}_3=1.10\times 10^{-12},\nonumber\\
&&{\cal B}_{12}=-1.26\times 10^{-9},\;\;\;{\cal B}_{13}=8.79\times 10^{-11},\;\;\; {\cal B}_{23}=-1.21\times 10^{-12}.\eeqn
Obviously, except ${\cal B}_1$, others are negligible since their sum is below the percent level relative to ${\cal B}_1$ (For instance, the ratio ${\cal B}_2/{\cal B}_1$ is about $0.2\%$). Likewise, one can find that only ${\cal B}_2$ will get a little smaller and others remain almost unchanged for the dimuon final state. This is consistent with the statement in Refs. \cite{Fleming93, Fleming94} that, contributions from Fig. \ref{figure2}, relative to the ones from Fig. \ref{figure1}, are suppressed by the factor $m_V^2/m_Z^2$, which, for $J/\Psi$, is below the percent level or even smaller (In calculating ${\cal M}_1$, the contribution through $Z\to \ell^+\ell^- Z^*$, followed by $Z^*\to V$, has been included. Actually, this part can be neglected for the charmonium final states). Here we wish to emphasize that, to calculate the amplitude of Fig. \ref{figure2}, one has to introduce the non-perturbative quantity $\psi_V(0)$, which may not be determined very precisely; on the other hand, the non-perturbative effects due to strong interactions from Fig. \ref{figure1}, are confined to the matrix element (\ref{fv}) up to the decay constant $f_V$. As pointed out in Ref. \cite{BBLY06}, this approach has the advantage that it automatically takes into account higher order corrections to the electromagnetic current, which are common to both the electromagnetic decay and production of a vector meson. Since $f_{J/\Psi}$ can be fixed from the measured $J/\Psi\to e^+ e^-$ width, a clean prediction for $Z\to J/\Psi\ell^+\ell^-$ with $\ell=e,~\mu$ can be expected. We thus obtain
\beq\label{brJPsitot}
{\cal B}(Z\to J/\Psi \ell^+\ell^-)=(7.78\pm0.14)\times 10^{-7},
\eeq
where the error is due to the uncertainty of the experimental value of $\Gamma(J/\Psi\to e^+e^-)$.

Similarly, for another $1^{--}$ charmonium $\Psi(2S)$, it is enough to consider the contribution from Fig. \ref{figure1}. Numerically, taking $\Gamma(\Psi(2S)\to e^+e^-)=(2.33\pm 0.04)$ keV \cite{PDG24}, we have
\beq\label{brPsitot}
{\cal B}(Z\to \Psi(2S) \ell^+\ell^-)=(2.40\pm0.04)\times 10^{-7}.
\eeq

Now for the charmonium case, the decay rate of $Z\to V\ell^+\ell^-$ can be written as
\beq\label{rateofcharm}
\Gamma(Z\to V\ell^+\ell^-)=\frac{ g_Z^2[({g}^\ell_V)^2+({g}^\ell_A)^2]}{32\pi^2 m_Z^2}\frac{\Gamma(V\to e^+ e^-)}{m_V}~I(m_V^2)\eeq
with
\beqn\label{ImV}
I(m_V^2)=\int^{(m_Z-m_V)^2}_{4m_\ell^2}ds~\int^1_{-1}d\cos\theta~\beta_\ell~\lambda^{1/2}(m_Z^2, m_V^2,s)~ \tilde{I}(s, \cos\theta),
\eeqn
where the form of $\tilde{I}(s, \cos\theta)$ is explicitly shown in eq. (\ref{tildeI}) of the Appendix. Therefore, the ratio of ${\cal B}(Z\to J/\Psi\ell^+\ell^-)$ to ${\cal B}(Z\to \Psi(2S) \ell^+\ell^-)$ can be determined up to the ratio of the experimental width of the charmonium leptonic decays, which reads
\beq\label{ratioJPsiPsi}
R=\frac{{\cal B}(Z\to J/\Psi\ell^+\ell^-)}{{\cal B}(Z\to \Psi(2S) \ell^+\ell^-)}=\frac{m_{\Psi(2S)}}{m_{J/\Psi}}\frac{I(m_{J/\Psi}^2)}{I(m_{\Psi(2S)}^2)}\frac{\Gamma(J/\Psi\to e^+e^-)}{\Gamma(\Psi(2S)\to e^+e^-)}.\eeq
Using the current experimental data, we have $R=3.24\pm 0.08$. Thus, the central value of ${\cal R}_{J/\Psi\ell^+\ell^-}$ given by the CMS Collaboration \cite{CMS18}, shown in eq. (\ref{RofZtoJpsi}), in which $R=3.5$ was used, will be slightly shifted as ${\cal R}_{J/\Psi\ell^+\ell^-}=0.65$.

Next, let us deal with the processes containing the bottomonium final state. By taking $\Gamma(\Upsilon(1S)\to e^+e^-)=1.340\pm 0.018$ keV \cite{PDG24} and $\psi^2_{\Upsilon(1S)}=0.512\pm 0.035$ GeV$^3$ from Ref. \cite{CLY11}, we obtain
\beqn\label{brUpsilon1S}
&&{\cal B}_1=2.04\times 10^{-8},\;\;\; {\cal B}_2=9.29\times 10^{-10},\;\;\; {\cal B}_3=8.77\times 10^{-12},\nonumber\\
&&{\cal B}_{12}=4.66\times 10^{-10},\;\;\;{\cal B}_{13}=-7.21 \times 10^{-11},\;\;\; {\cal B}_{23}=-2.61\times 10^{-12}\eeqn
for $Z\to \Upsilon(1S) e^+ e^-$; while for the muon mode, only ${\cal B}_2$ will be changed as $4.29\times 10^{-10}$.  One can easily find that ${\cal B}_1$ is still the most dominant, however, the other contributions from Fig. \ref{figure2} via $Z\to V\gamma^*\to V\ell^+\ell^-$ can reach the percent level, and will be non-negligible any more (For instance, the ratio ${\cal B}_2/{\cal B}_1$ is about $5\%$). Meanwhile, the contributions via the virtual $Z$ transition remain insignificant. Totally, we have
\beqn\label{brUpsilontot-ee}
{\cal B}(Z\to \Upsilon(1S) e^+ e^-)=(2.18\pm 0.03)\times 10^{-8},\\\label{brUpsilontot-mumu}
{\cal B}(Z\to \Upsilon(1S) \mu^+ \mu^-)=(2.13\pm 0.03)\times 10^{-8},
\eeqn
where the error is almost entirely from the uncertainty of the measured $\Upsilon(1S)\to e^+e^-$ decay width, and the error from the uncertainty of $\psi^2_{\Upsilon(1S)}(0)$ can be neglected. Note that, relative to the dominant fragmentation contribution, the inclusion of other contributions changes the branching fraction by $+6\%$ for $Z\to \Upsilon(1S) e^+ e^-$, and by $+4\%$ for $Z\to \Upsilon(1S)\mu^+\mu^-$. For comparison,  ${\cal B}(Z\to \Upsilon(1S)\ell^+\ell^-)=2\times 10^{-8}$ was predicted in Ref. \cite{Fleming93}.

In the present work, in order to evaluate the diagrams in Fig. \ref{figure2}, we perform the computation in the framework of the nonrelativistic color-singlet model. Here we would like to give some remarks on the uncertainties of our theoretical predictions from this approach.

First, the non-perturbative quantity $\psi_V(0)$, which is taken as an input, cannot be determined precisely. Its uncertainty has been included in our numerical calculation, whose effects are negligible for the bottomonium final states. As to the charmonium case, the contribution from Fig. \ref{figure2} can be neglected.

Second, one may go beyond the present approach by using the nonrelativistic QCD (NRQCD) factorization method \cite{BBL95} to perform a systematical calculation in powers of $\alpha_s$ ($\alpha_s=g_s^2/4\pi$ and $g_s$ is the strong coupling constant) and $v$ ($v$ is the heavy-quark velocity in the quarkonium rest frame).  Actually, our present computation based on the color-singlet model are equivalent to the one based on the NRQCD approach at the leading order. Therefore, to estimate the theoretical errors of our results, one can assume that the uncalculated QCD corrections in $\alpha_s$ are of order $\alpha_s(m_V)$ and that the uncalculated corrections in $v$ are of order $v^2$.  On the other hand, the calculation of ${\cal M}_1$ (then to $\Gamma_1$), which hadronizes the electromagnetic currents based on eq. (\ref{fv}), already contains the QCD and relativistic corrections to all orders. Thus the estimation of theoretical uncertainties from uncalculated corrections of order $\alpha_s$ and $v^2$ will be applied to $\Gamma_i$'s with $i\neq 1$. In the following, one will take
\beq\label{asv2-charm}
\alpha_s(m_V) = 0.25,\;\;\;  v^2 = 0.3
\eeq
for charmonium, and
\beq\label{asv2-bottom}
\alpha_s(m_V) = 0.18,\;\;\; v^2 = 0.1
\eeq
for bottomonium, respectively, to carry out the analysis.

Since the contributions from Fig. \ref{figure2} can be neglected for charmonium final states, the theoretical uncertainties from $\alpha_s$ and $v$ for ${\cal B}(Z\to J/\Psi\ell^+\ell^-)$ and ${\cal B}(Z\to \Psi(2S)\ell^+\ell^-)$ will be vanishingly small, far below the uncertainties due to the measured $\Gamma(V\to e^+e^-)$. In the case of $\Upsilon(1S)$, using eq. (\ref{asv2-bottom}), one can obtain the theoretical error is about $(\pm 0.03)\times 10^{-8}$ for ${\cal B}(Z\to \Upsilon(1S)e^+ e^-)$, and $(\pm 0.02)\times 10^{-8}$ for ${\cal B}(Z\to \Upsilon(1S)\mu^+ \mu^-)$, respectively,  which are comparable to the errors in eqs. (\ref{brUpsilontot-ee}) and (\ref{brUpsilontot-mumu}). Therefore, we have
\beqn\label{brUpsilontot-th-error-ee}
{\cal B}(Z\to \Upsilon(1S) e^+e^-)=(2.18\pm 0.03\pm 0.03)\times 10^{-8}=(2.18\pm 0.04)\times 10^{-8},\\\label{brUpsilontot-th-error-mumu}
{\cal B}(Z\to \Upsilon(1S) \mu^+\mu^-)=(2.13\pm 0.03\pm 0.02)\times 10^{-8}=(2.13\pm 0.04)\times 10^{-8}.
\eeqn

Similarly, using $\Gamma(\Upsilon(2S)\to e^+e^-)=0.612\pm 0.011$ keV and $\Gamma(\Upsilon(3S)\to e^+e^-)=0.443\pm 0.008$ keV \cite{PDG24}, together with $\psi^2_{\Upsilon(2S)}=0.271\pm 0.019$ GeV$^3$ and $\psi^2_{\Upsilon(3S)}=0.213\pm 0.015$ GeV$^3$ from Ref. \cite{CLY11}, we obtain
\beqn\label{brUpsilon2S3Stot}
{\cal B}(Z\to \Upsilon(2S) e^+ e^-)=(0.88\pm 0.02)\times 10^{-8},\\ {\cal B}(Z\to \Upsilon(2S) \mu^+ \mu^-)=(0.86\pm 0.02)\times 10^{-8},
\\
{\cal B}(Z\to \Upsilon(3S) e^+ e^-)=(0.60\pm 0.01)\times 10^{-8}, \\ {\cal B}(Z\to \Upsilon(3S) \mu^+ \mu^-)=(0.58\pm 0.01)\times 10^{-8}.
\eeqn
Our calculation shows that the inclusion of Fig. \ref{figure2} changes the branching fraction by $+8\%$ for $Z\to \Upsilon(2S) e^+e^-$, by $+5\%$ for $Z\to \Upsilon(2S) \mu^+\mu^-$, by $+9\%$ for $Z\to \Upsilon(3S) e^+ e^-$, and by $+5\%$ for $Z\to \Upsilon(3S) \mu^+ \mu^-$.

\begin{figure}[t]
\begin{center}
\includegraphics[width=8cm,height=5cm]{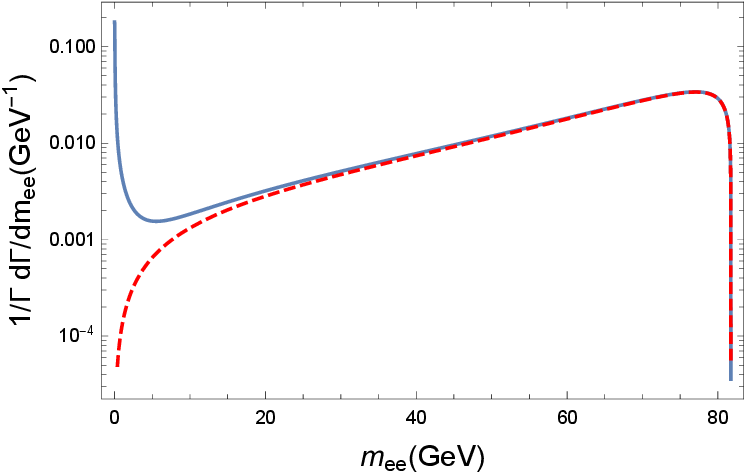}
\end{center}
\caption{The normalized invariant mass distribution of $Z\to\Upsilon(1S) e^+ e^-$ decay, where $m_{ee}=\sqrt{s}$ is the dilepton invariant mass. The red-dashed line denotes the contribution from Fig. 1 only while the solid line gives the total contribution.}\label{figure3}
\end{figure}

Besides the branching ratios, one can further explore the differential distributions of $Z\to V\ell^+\ell^-$ decays. The invariant mass distributions for the lepton pair or the energy distributions for the vector mesons have been studied in Refs. \cite{BR90, Fleming93} in which only Fig. \ref{figure1} was considered.  For the charmonium case, since the contribution generated from Fig. \ref{figure2} can be neglected, these differential distributions should of course remain almost unchanged.

For the bottomonium case, let us take $Z\to \Upsilon(1S) \ell^+\ell^-$ as an example. After integrating over $\cos\theta$ in eq. (\ref{distribution1}), the normalized distribution of the electron mode with respect to the dilepton invariant mass $m_{ee}=\sqrt{s}$ has been plotted in Fig. \ref{figure3}. For comparison, we also show the decay spectrum from Fig. \ref{figure1} only, which has been studied in Ref. \cite{BR90}.  As can be seen, in the most range of $m_{ee}$, especially in the large dilepton invariant mass region, the inclusion of Fig. \ref{figure2} will not bring about any significant effects. However, in the low $m_{ee}$ region, below about 15 GeV, the interesting deviation could be found. The peak in the plot, giving the dominant contribution for small $m_{ee}$, is obviously due to the virtual photon pole induced by $Z\to V\gamma^*\to V e^+ e^-$ transition. We are not going to display the plot for the dimuon invariant mass distribution of $Z\to \Upsilon(1S) \mu^+ \mu^-$ since it is believed that one will achieve the similar behavior as the above figure. From the plot in Fig. \ref{figure3}, it is easy to understand that, due to the smallness of the electron mass, the electron mode could have the relative larger branching fraction than the muon mode because of the virtual photon contribution from Fig. \ref{figure2}.

\begin{figure}[t]
\begin{center}
\includegraphics[width=6.5cm,height=4.5cm]{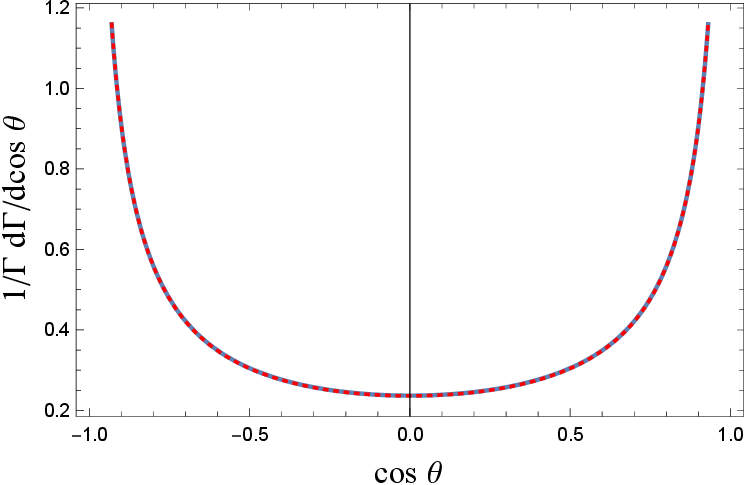}\hspace{0.25in}
\includegraphics[width=6.5cm,height=4.5cm]{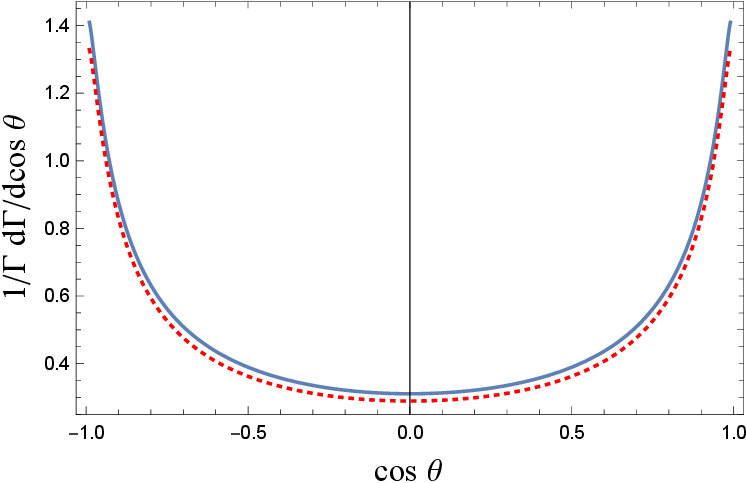}
\end{center}
\caption{The normalized angular distributions of $Z\to J/\Psi \ell^+ \ell^-$ (left plot) and $Z\to \Upsilon(1S)\ell^+\ell^-$ (right plot) decays. The red-dashed line denotes the contribution from Fig. 1 only. }\label{figure4}
\end{figure}

On the other hand, by integrating over $s$ in eq. (\ref{distribution1}), one can also examine the differential angular distributions of $Z\to V\ell^+\ell^-$ decays. The resulting plots for $Z\to J/\Psi\ell^+\ell^-$  and $Z\to \Upsilon (1S)\ell^+\ell^-$ have been given in Fig. \ref{figure4}, respectively. Similar to the case of the invariant mass distribution, the inclusion of Fig. \ref{figure2} leads to a very tiny shift for the bottomomuim mode; while, for the charmonium process, the contribution of Fig. \ref{figure2} is basically no visible effect. One can easily find that the angular distribution is symmetric under $\cos\theta \leftrightarrow -\cos\theta$. This indicates that forward-backward asymmetries for the final leptons in $Z\to V\ell^+\ell^-$ would be zero in the SM.

\subsection{Beyond the SM}

Our analysis above shows that no forward-backward asymmetry in $Z\to V\ell^+\ell^-$ can be expected if considering these processes in the SM only. On the other hand, the nonzero forward-backward asymmetry in these decays means that CP violation occurs in $Z$-boson decays, which thus could be an interesting physical observable to search for new physics beyond the SM.

\begin{figure}[t]
\begin{center}
\includegraphics[width=6cm,height=3cm]{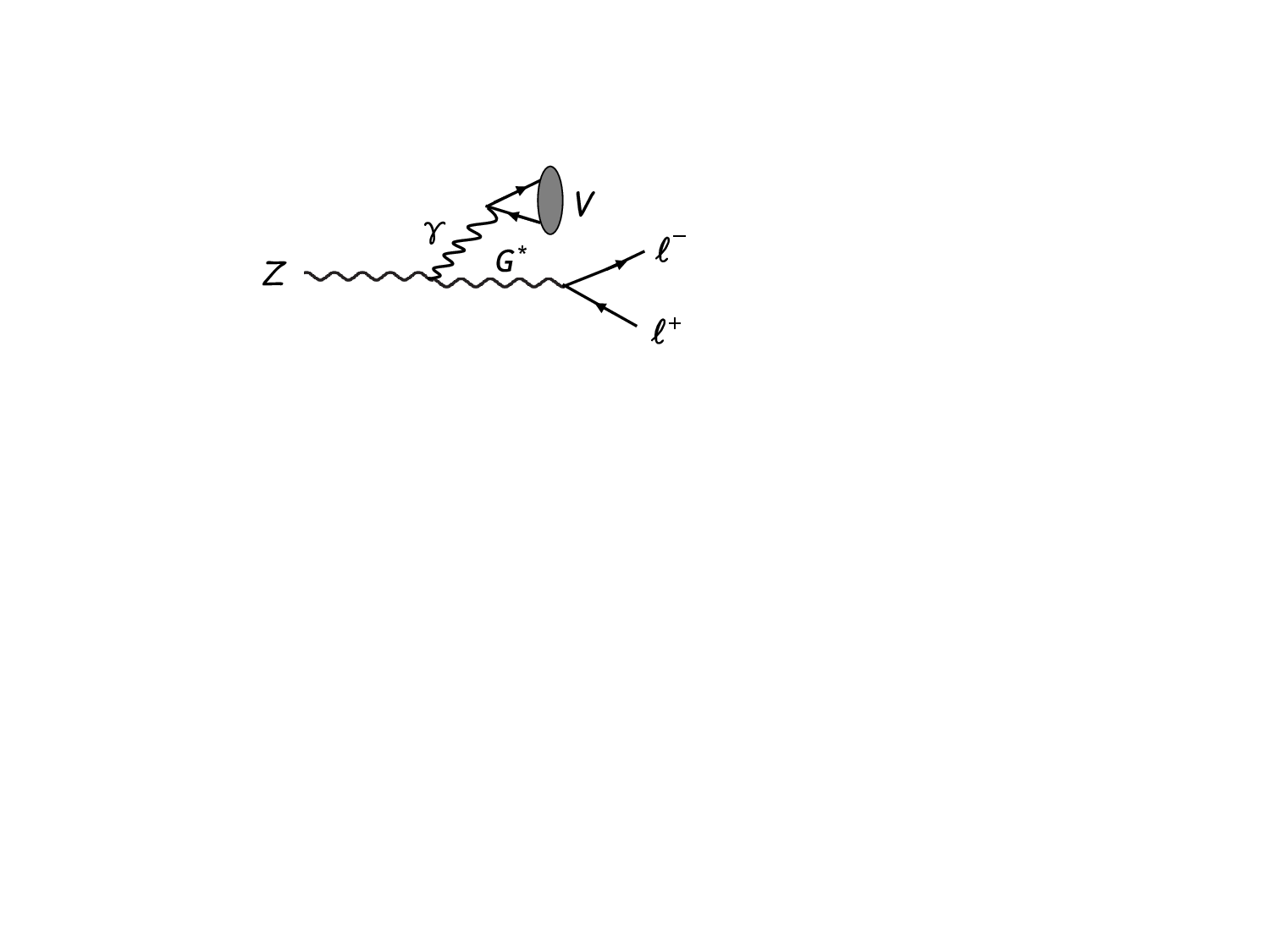}
\end{center}
\caption{The Feynman diagram contributing to CP-violating amplitude in $Z\to V\ell^+\ell^-$ decays by the effective $Z\gamma G^*$ vertex.}\label{figure5}
\end{figure}

It is known that anomalous neutral triple gauge couplings (NTGC)\cite{HPZ87, BB93}, which are absent in the SM, could induce CP-violating effects in $Z$-boson decays \cite{LPTT01, PR05}. The general form of the CP-violating NTGC involving the effective $Z(p)\gamma (q) G^*(k)$ vertex with $G=Z,~\gamma$ denoting the virtual gauge boson, has been given in Refs. \cite{HPZ87, BB93, GLR00}, which reads
\beqn\label{CPVNTGC}
\Gamma_{\alpha\beta\mu}^{Z\gamma G^*}(p,q,k)=\frac{ie}{m_Z^2}\left[ h_1^G(q_\alpha g_{\mu\beta}-q_\beta g_{\mu\alpha})+\frac{h_2^G}{m_Z^2}k_\alpha(k\cdot q g_{\mu\beta}-q_\beta k_\mu)\right](k^2-m_G^2),
\eeqn
where $h_i^G$'s are effective parameters, which have been already bounded experimentally \cite{PDG24}. It is easy to check that the effective $\Gamma_{\alpha\beta\mu}^{Z\gamma G^*}$ can lead to the CP-violating amplitude of $Z\to V\ell^+\ell^-$ through $\gamma\to V$ and $G^*\to \ell^+\ell^-$, which has been displayed in Fig. \ref{figure5}. Obviously, $\gamma\to V$ transition will generate the factor $1/m_V^2$, which may enhance the amplitude. By comparison, the transition $G^*\to V$ with $\gamma\to \ell^+\ell^-$ will give the suppressed amplitude, therefore we do not include it.
Very recently, it has been claimed by the authors of Refs. \cite{EHX} that the above conventional CP-violating NTGC is incompatible with the spontaneous breaking of the full electroweak gauge group, and a different structure for the second dimension-8 term of eq. (\ref{CPVNTGC}) has been obtained in their work. In the present paper, we only consider the first dimension-6 $h_1^G$ term to carry out our analysis of the CP-violating effects in $Z\to V \ell^+\ell^-$ decays.\footnote{Generally speaking, the second dimension-8 term would be more suppressed by the high scale of new physics. However, its contribution might be important in some new scenarios. In the present work, for simplicity, we just choose the dimension-6 term to illustrate our numerical analysis.} Therefore, the new CP-violating amplitude can be straightforwardly written as
\beqn\label{amp-CPV}
i{\cal M}_{\rm new}=\frac{ie^2g_Z f_V h_1^Z}{2 m_Z^2m_V}\epsilon_\alpha(p)\epsilon_\mu^*(q) (q^\alpha g^{\mu\beta}-q^\beta g^{\mu\alpha})\bar{u}(k_1)\gamma_\beta(g_V^\ell-g_A^\ell\gamma_5)v(k_2)\nonumber\\
-\frac{ie^3 Q_Vf_V h_1^\gamma}{m_Z^2m_V}\epsilon_\alpha(p)\epsilon_\mu^*(q) (q^\alpha g^{\mu\beta}-q^\beta g^{\mu\alpha})\bar{u}(k_1)\gamma_\beta v(k_2),
\eeqn
where the first term is given by the anomalous $ZZ^*\gamma$ vertex, and the second one is by the $Z\gamma^*\gamma$ vertex. Thus, the interference between this new amplitude and the SM amplitude (\ref{amp-tot}) will generate the CP-violating forward-backward asymmetry in $Z\to V\ell^+\ell^-$ decays.

\begin{figure}[t]
\begin{center}
\includegraphics[width=8cm,height=5cm]{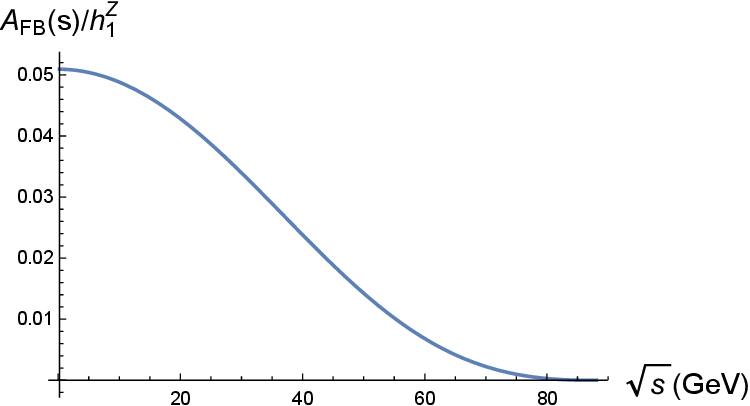}
\end{center}
\caption{The forward-backward asymmetry $A_{\rm FB}(s)/h_1^Z$ as a function of $\sqrt{s}$ in $Z\to J/\Psi \ell^+\ell^-$ decays.}\label{figure6}
\end{figure}

Note that, for the charmonium process $Z\to J/\Psi \ell^+\ell^-$, it is enough to include the contribution to its SM amplitude from Fig. \ref{figure1} only, namely, ${\cal M}_1^\gamma$ in eq. (\ref{amp0}). Direct calculation shows that, the interference of the first term of eq. (\ref{amp-CPV}) and ${\cal M}_1^\gamma$ is proportional to $h_1^Z [(g_V^\ell)^2+(g_A^\ell)^2]$; while for the second term of (\ref{amp-CPV}), the interference is proportional to $h_1^\gamma g_V^\ell$. Experimentally, the effective couplings have been bounded as \cite{Schael13}
\beqn\label{h1}
-0.12 < h_1^Z < +0.11,\;\;\; -0.05 < h_1^\gamma < +0.05,
\eeqn
and for the charged lepton, $g_V^\ell=-0.04$ and $g_A^\ell=0.5$, this seems to indicate that the contribution via the $ZZ^*\gamma$ vertex may give rise to more significant CP-violating effects in these decays. In order to illustrate our numerical study, in the following, we focus on the forward-backward asymmetry in $Z\to J/\Psi \ell^+\ell^-$ induced by the anomalous $ZZ^*\gamma$ vertex. Thus, using eq. (\ref{AFB1}), we obtain
\beq\label{AFB2}
A_{\rm FB}(s)=4 h_1^Z \cdot \frac{{\cal A}_1(s)}{{\cal A}(s)},
\eeq
where
\beqn\label{A1s}
{\cal A}_1(s)=\int^1_0 d\cos\theta ~\beta_\ell~\lambda^{1/2}(m_Z^2, m_{J/\Psi}^2, s)\frac{q\cdot k_1 q\cdot k_2-m_{J/\Psi}^2 k_1\cdot k_2}{(m_{J/\Psi}^2+2 k_1\cdot q)(m_{J/\Psi}^2+2 k_2\cdot q)}\frac{q\cdot(k_2-k_1)}{m_Z^2}\nonumber\\
-\int^0_{-1} d\cos\theta~ \beta_\ell~\lambda^{1/2}(m_Z^2, m_{J/\Psi}^2, s)\frac{q\cdot k_1 q\cdot k_2-m_{J/\Psi}^2 k_1\cdot k_2}{(m_{J/\Psi}^2+2 k_1\cdot q)(m_{J/\Psi}^2+2 k_2\cdot q)}\frac{q\cdot(k_2-k_1)}{m_Z^2},
\eeqn
and
\beqn\label{As}
{\cal A}(s)=\int^1_{-1}d\cos\theta~\beta_\ell~\lambda^{1/2}(m_Z^2, m_{J/\Psi}^2,s)~\tilde{I}(s, \cos\theta).
\eeqn
The expression of $\tilde{I}(s, \cos\theta)$ can be found in eq. (\ref{tildeI}) of the Appendix in which $m_V^2$ should be taken as $m_{J/\Psi}^2$.
Here we do not consider the contribution from the $h_1^Z$ term to ${\cal A}(s)$ since it is ${\cal O}[(h_1^Z)^2]$ and negligible for the present range of $h_1^Z$ in (\ref{h1}).  Obviously, using eqs. (\ref{qk1}) and (\ref{qk2}), the term $q\cdot (k_2-k_1)$ in (\ref{A1s}) can give rise to the linear term of $\cos\theta$. We have plotted $A_{\rm FB}(s)/h_1^Z$ with respect to $\sqrt{s}$, the dilepton invariant mass, in Fig. \ref{figure6}. Using the current bound on $h_1^Z$, the forward-backward asymmetry in $Z\to J/\Psi \ell^+\ell^-$, given by the anomalous $ZZ^*\gamma$ vertex, could be up to $10^{-3}$.

Meanwhile, one can further obtain the integrated forward-backward asymmetry by integrating over $s$, which is expressed as
\beq\label{AFB3}
{\cal A}_{\rm FB}=4 h_1^Z\cdot \Frac{\int^{s_{\rm max}}_{s_{\rm min}} d s ~{\cal A}_1(s)}{\int^{s_{\rm max}}_{s_{\rm min}} d s ~{\cal A}(s)},\eeq
where $(s_{\rm min}, s_{\rm max})$ denotes the range of the integration over $s$. For the full phase space, i.e. $4m_\ell^2 < s < (m_Z-m_{J/\Psi})^2$, we have
\beq\label{integratedAFB1}
{\cal A}_{\rm FB}=0.0056 h_1^Z.
\eeq
From the current bound of $h_1^Z$, this integrated asymmetry could be $6\times 10^{-4}$, which is quite small and will be challenging experimentally. On the other hand, when we calculate the asymmetry by imposing some cut on $s$, ${\cal A}_{\rm FB}$ may be enhanced. If taking $(5{\rm GeV})^2 < s < (60 {\rm GeV})^2$ by cutting the large $s$ region, we find
\beq\label{integratedAFB2}
{\cal A}_{\rm FB}=0.021 h_1^Z,
\eeq
which means the integrated asymmetry could reach the ${\cal O}(10^{-3})$ level.\footnote{The decay rate of this process is dominated by the electromagnetic fragmentation transition, which is in the large $s$ region. Thus, in the above cut region, the denominator of eq. (\ref{AFB3}) will be substantially decreased, which can enhance the integrated asymmetry.} However, as we know, these decays are dominated by the electromagnetic fragmentation transition, the most events, as shown in Fig. \ref{figure3}, will intend to be in the large $s$ region. Thus, it is still not easy for experimental observations.

One can extend our analysis to the case of other heavy quarkonium final states $\Psi(2S)$, $\Upsilon(nS)$ with $n=1, 2, 3$ etc. In the present paper, in order to illustrate our numerical study, we have assumed that possible CP-violating effects in $Z\to V \ell^+\ell^-$ decays are induced by the anomalous NTGC only. This is not a necessary assumption. For example, the general structure of CP-violation in $Z\to \gamma\ell^+\ell^-$ decays has been studied in the framework of the effective Lagrangian formalism by the authors of Ref. \cite{BLMN89}. Certainly, this formalism can be applied in the present analysis, which will be left for a future separate publication.

\section{Summary and outlook}

We have investigated exclusive rare $Z$-boson decays into a heavy vector quarkonium plus a lepton pair $\ell^+\ell^-$ with $\ell=e,~\mu$. It is assumed in the past literature that the leading order SM contribution to these modes comes from $Z\to \gamma^*\ell^+\ell^-$ with the subsequent fragmentation transition $\gamma^*\to V$, and the other contributions such as $Z\to V\gamma^*\to V\ell^+\ell^-$ are strongly suppressed, which do not need to be calculated. In order to provide up-to-date theoretical predictions for these decays for use in future high statistics experimental machines, we have presented a systematical analysis of the tree-level SM contributions to these rare modes by including all of the relevant Feynman diagrams, which have been displayed in Fig. \ref{figure1} and Fig. \ref{figure2}, respectively. Our calculation demonstrates that, for the processes containing the charmonium final states, the diagram via the electromagnetic fragmentation transition almost takes control of the whole thing and the rest part is indeed negligible; while for $Z\to \Upsilon(nS)\ell^+\ell^-$ decays, the full calculation of the tree-level diagrams in the SM increases branching fractions of these processes by around $4\%\sim 9\%$, relative to the ones via the electromagnetic fragmentation transition in Fig. \ref{figure1} only. Therefore, the calculation of $Z\to J/\Psi \ell^+\ell^-$ and $Z\to \Psi(2S)\ell^+\ell^-$ decay rates could be, in a large extent, free of the contamination of nonperturbative QCD effects, and rather clean predictions have been obtained. On the other hand, in the bottomonium processes the inclusion of Fig. \ref{figure2} may give rise to some additional theoretical errors from the uncalculated QCD corrections and relativistic corrections, which have been estimated and show that theoretical uncertainties of ${\cal B}(Z\to \Upsilon(nS)\ell^+\ell^-)$ are well under control.

We have also analyzed the differential distributions of these decays, as displayed in Fig. \ref{figure3} and Fig. \ref{figure4}, respectively. It is very natural that both dilepton invariant mass and angular distributions are dominated by the contributions from Fig. \ref{figure1}. For the dilepton invariant mass distribution of the channels containing bottomonium final states, a peak appears in the low $\sqrt{s}$ region due to the virtual photon pole induced by $Z\to V\gamma^*\to V \ell^+ \ell^-$ transition.
From the angular distributions, It is easy to find that there are no CP-violating forward-backward asymmetries for these rare processes in the SM.  One can however obtain the CP-violating amplitude by introducing the anomalous NTGC terms, whose effective parameters could be bounded experimentally. The explicit analysis shows that, for $Z\to J/\Psi \ell^+\ell^-$, the present experimental constrains allow the integrated forward-backward asymmetry to be up to $6\times 10^{-4}$ and the differential $A_{\rm FB}(s)$ to reach $10^{-3}$. Thus it is expected that high-precision measurements of these CP-violating observables might probe interesting extensions of the SM or impose significant bounds on the effective NTGC paramters.

It has been concerned by experimental particle physicists to search for exclusive rare weak gauge bosons processes containing hadronic final states. However, so far no evidence for such decays has been obtained experimentally. For rare $Z$-boson decays, the analysis of $Z\to V\ell^+\ell^-$ in the past literature and the present paper seems to indicate these three-body $Z$ decays may be the promising candidates in future experiments, especially in some facilities with huge $Z$ events accumelated. Since theoretical predictions on $Z\to V\ell^+\ell^-$ decays in the SM can be rather clean, experimental studies of them would help both to test the SM and to look for new physics scenarios.

The enormous samples of $Z$ bosons will become available in the high-luminosity LHC \cite{HLLHC19}, or other future experimental machines such as CEPC \cite{CEPC} and FCC-ee \cite{FCC-ee}. In particular, these two lepton machines will be running at the $Z$ mass region for a period time, thus a huge number of events, e.g., about $6\cdot 10^{12}$ $Z$ bosons would be produced at FCC-ee. We are eagerly looking forward to some dedicated studies for exploring rare $Z$-boson decays such as $Z\to V\ell^+\ell^-$ being performed at these facilities.

\section*{Acknowledgments}
This work was supported in part by the National Natural Science Foundation of China under Grants No. 12247103 and No. 12547106.

\appendix
\newcounter{pla}
\renewcommand{\thesection}{\Alph{pla}}
\renewcommand{\theequation}{\Alph{pla}\arabic{equation}}
\setcounter{pla}{1}
\setcounter{equation}{0}

\section*{Appendix: Some analytic expressions of the differential decay rate}

The full expression of $Z\to V\ell^+\ell^-$ decay at the tree-level in the SM can be derived using eq. (\ref{distribution1})  together with eqs. (\ref{amp2}), (\ref{amp1}), (\ref{amp3-Z}) and ({\ref{amp-tot}}).  Here we take ${\Gamma_1}$ as an example to show our derivation explicitly.
\beqn\label{distribution-gamma1}
\frac{d^2\Gamma_1}{d s ~d\cos\theta}&=&\frac{1}{512\pi^3m_Z^3}\beta_\ell ~\lambda^{1/2}(m_Z^2, m_V^2,s)~ \frac{1}{3}\sum_{\rm spins}|{\cal M}_1|^2\nonumber\\
&=&\frac{g_Z^2\Gamma(V\to e^+e^-)}{32\pi^2m_Z^2 m_V}[(\tilde{g}_V^\ell)^2+(\tilde{g}_A^\ell)^2]\beta_\ell \lambda^{1/2}(m_Z^2,m_V^2,s)\tilde{I}(s, \cos\theta)
\eeqn
with
\beqn\label{tildeI}
\tilde{I}(s, \cos\theta)=\left[\frac{2 p\cdot k_1 p\cdot k_2-m_Z^2 k_1\cdot k_2}{(m_{V}^2+2 k_1\cdot q)^2}+\frac{2 p\cdot k_1 p\cdot k_2-m_Z^2 k_1\cdot k_2}{(m_{V}^2+2 k_2\cdot q)^2}\right.\nonumber\\\left. +\frac{2k_1\cdot k_2(m_Z^2+m_{V}^2)}{(m_{V}^2+2 k_1\cdot q)(m_{V}^2+2 k_2\cdot q)}\right].
\eeqn
On the other hand, one can easily derive
\beqn
p\cdot k_1=\frac{m_Z^2-m_V^2+s}{4}-\frac{1}{4}\beta_\ell \lambda^{1/2}(m_Z^2, m_V^2, s)\cos\theta,\label{pk1}\\
p\cdot k_2=\frac{m_Z^2-m_V^2+s}{4}+\frac{1}{4}\beta_\ell \lambda^{1/2}(m_Z^2, m_V^2, s)\cos\theta,\label{pk2}\\
m_V^2+2 k_1\cdot q=\frac{m_Z^2+m_V^2-s}{2}-\frac{1}{2}\beta_\ell \lambda^{1/2}(m_Z^2,m_V^2,s)\cos\theta,\label{qk1}\\
m_V^2+2 k_2\cdot q=\frac{m_Z^2+m_V^2-s}{2}+\frac{1}{2}\beta_\ell \lambda^{1/2}(m_Z^2,m_V^2,s)\cos\theta. \label{qk2}
\eeqn
By further neglecting the lepton mass, we have $k_1\cdot k_2=s/2$. This shows that the function $\tilde{I}$ and the differential rate (\ref{distribution-gamma1}) can be in terms of the kinematical variables $s$ and $\cos\theta$ completely. It is straightforward to see that this differential rate is the even function of $\cos\theta$.  One can similarly deal with the rest parts in eq. (\ref{distribution1}).

\end{document}